\documentclass[aip,twocolumn,showpacs,superscriptaddress,reprint,amsmath,amssymb,floatfix]{revtex4-1}
\pdfoutput=1
\usepackage{graphicx}
\usepackage{dcolumn}
\usepackage{bm}
\usepackage{amssymb}
\usepackage{epsfig}
\usepackage{pdfsync}
\usepackage[colorlinks,linkcolor=blue,citecolor=blue,urlcolor=blue]{hyperref}

\begin{document}
\title{Proximity nanovalve with large phase-tunable thermal conductance}
\author{E. Strambini}
\email{e.strambini@sns.it}
\affiliation{NEST Istituto Nanoscienze-CNR  and Scuola Normale Superiore, I-56127 Pisa, Italy}
\author{F. S. Bergeret}
\email{sebastian\_bergeret@ehu.es}
\affiliation{Centro de F\'{i}sica de Materiales (CFM-MPC), Centro Mixto CSIC-UPV/EHU, Manuel de Lardizabal 4, E-20018 San Sebasti\'{a}n, Spain}
\affiliation{Donostia International Physics Center (DIPC), Manuel de Lardizabal 5, E-20018 San Sebasti\'{a}n, Spain}
\affiliation{Institut f\"ur Physik, Carl von Ossietzky Universit\"at, D-26111 Oldenburg, Germany}
\author{F. Giazotto}
\email{f.giazotto@sns.it}
\affiliation{NEST Istituto Nanoscienze-CNR  and Scuola Normale Superiore, I-56127 Pisa, Italy}
\begin{abstract}
We propose a phase-controlled heat-flux quantum valve based on the proximity effect driven by a superconducting quantum interference proximity transistor (SQUIPT).
Its operation relies on the phase-dependent quasiparticle density of states in the Josephson weak-link of the SQUIPT which controls  thermal transport across the device.
In a realistic Al/Cu-based setup the structure can provide efficient control of thermal current inducing temperature swings exceeding $\sim100$~mK, and flux-to-temperature transfer coefficients up to $\sim 500$~mK/$\Phi_0$ below 100~mK. 
The nanovalve performances improve by lowering the bath temperature, making the proposed structure a promising building-block for the implementation of coherent caloritronic devices operating below 1~K.
\end{abstract}
\maketitle

Phase-dependent manipulation of heat in solid-state nanodevices is nowadays one of the major challenges of coherent caloritronics~\cite{Martinez-Perez_Coherent_2014}, and plays a key role in determining the physical properties of mesoscopic systems at low temperature. 
Toward this direction, the prototype for a heat interferometer has been recently realized with a superconducting quantum interference device (SQUID) where the modulation of the thermal current has been achieved thanks to the Josephson coupling~\cite{Giazotto_Josephson_2012, Giazotto_Phase-controlled_2012,Martinez-Perez_Fully_2013,Martinez-Perez_quantum_2014}. 
Yet, phase-dependent thermal transport has been also demonstrated in Andreev interferometers~\cite{Eom_Phase_1998, Dikin_Low-temperature_2001,Bezuglyi_Heat_2003} where the  proximity effect in a normal metal affects its thermal conductance, and is controlled via a magnetic field.

Here we propose an alternative approach to control heat transport by envisioning a thermal nanovalve based on proximity effect but phase-controlled by a SQUIPT~\cite{Giazotto_Superconducting_2010, Meschke_Tunnel_2011, Jabdaraghi_Non-hysteretic_2014, Ronzani_Highly-sensitive_2014}. 
Differing from SQUID-based and  Andreev interferometers our device allows a drastic quenching of the thermal conductance which makes our proposal an efficient phase-tunable thermal nanovalve.
Specifically, we expect an improvement of the temperature swing (up to $\sim 100$~mK) and a flux-to-temperature transfer function exceeding 500~mK$/\Phi_0$ at 100~mK.
\begin{figure}[th!]
\includegraphics[width=\columnwidth]{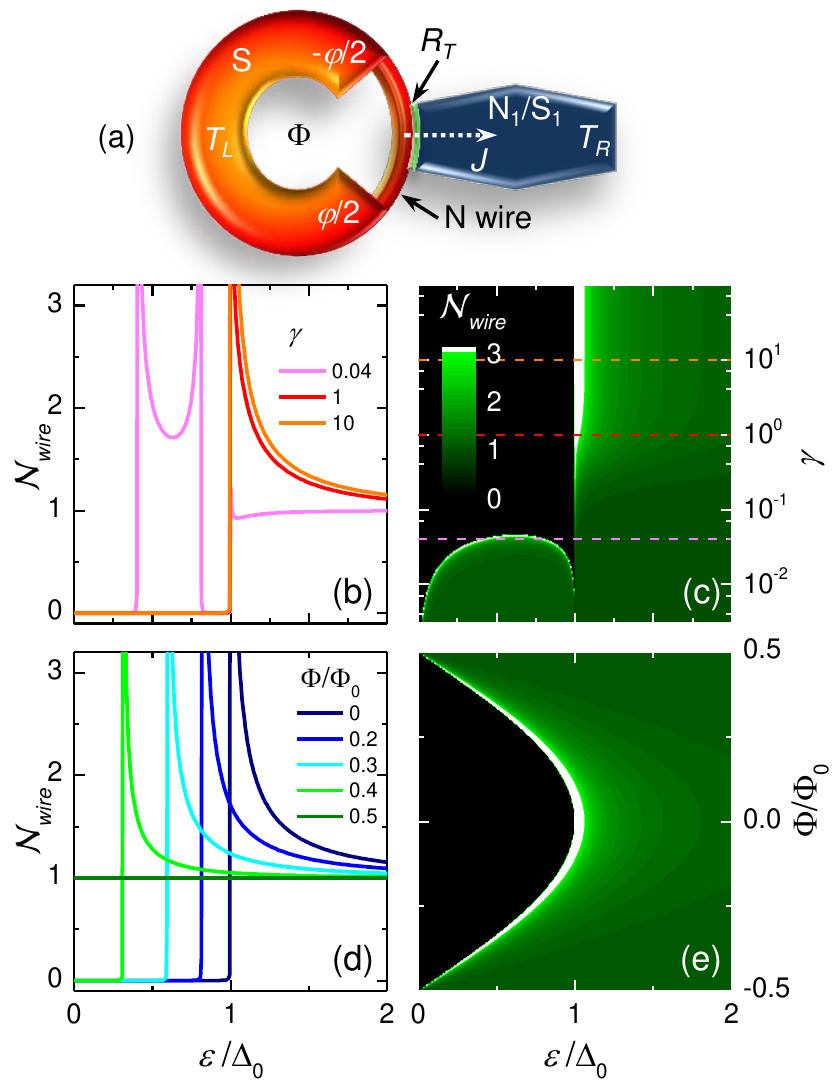}
\caption{\label{fig1} 
(a) Sketch of the  proximity nanovalve discussed in the text.
$J$ denotes the thermal current flowing through the structure. 
(b) $\mathcal{N}_{wire}$ vs energy $\varepsilon$ calculated for a few values of $\gamma$ at $\Phi=0$. 
(c) Color plot of $\mathcal{N}_{wire}$ vs $\varepsilon$ and $\gamma$ at $\Phi=0$. 
(d) $\mathcal{N}_{wire}$ vs $\varepsilon$ calculated for $\gamma= 10$ at different values of $\Phi$, and 
(e) corresponding color plot in $\varepsilon$ and $\Phi$. All the calculations were performed at zero temperature.
}
\end{figure}

A sketch for the proximity nanovalve is shown in Fig.~\ref{fig1}(a) and consists of a SQUIPT composed by a superconducting (S) ring interrupted by a diffusive normal metal (N) wire of length $L$. 
We assume the wire transverse dimensions to be negligible in comparison to its length so that it can be considered as quasi-one-dimensional.  Superconducting correlations are induced in the N wire owing to proximity effect from the S loop which modifies the wire density of states (DoS)~\cite{gueron_superconducting_1996}. 
In addition, a normal metal (N$_1$) or a superconducting electrode (S$_1$) identical to S  is tunnel-coupled to the middle of the wire through an insulating contact of negligible width  respect to the wire length. $R_T$ denotes the normal-state resistance of the junction. 
We assume the SN ring and the N$_1$(or S$_1$) electrode to be in steady-state thermal equilibrium and to reside at different temperatures $T_L$ and $T_R$, respectively, with $T_L\geq T_R$. 
The nanovalve is therefore only temperature biased.
In the limit of negligible geometric inductance of the ring it follows that $\varphi=2\pi\Phi/\Phi_0$, where $\varphi$ is the phase difference across the SN boundaries, $\Phi$ is the applied magnetic flux through the loop, and $\Phi_0=2.067\times 10^{-15}$ Wb is the flux quantum. 
The ring geometry allows to change $\varphi$ which leads to the modification of the N-wire DoS~\cite{Zhou_Density_1998} and, in turn, a drastic modification of \emph{thermal} transport through the device. 

The DoS in the  N  wire ($\mathcal{N}_{wire}$) is given by $\mathcal{N}_{wire}(\varepsilon,x)={|\rm Re}[g^R(\varepsilon,x)]|,$ 
where $ g_R$ is the normal retarded  quasiclassical Green's function,  $\varepsilon$ is the energy, and $x$ is the spatial coordinate along the wire.  The function $g^R$ can be obtained by solving the one-dimensional Usadel equation (we skip the supra index $R$)~\cite{Usadel_Generalized_1970} 
 \begin{equation}
\hbar D\partial_x(\hat g\partial_x\hat  g)+(\varepsilon+i\Gamma)\left[\hat \tau_z,\hat g\right]=0\label{Usadel}\; ,
\end{equation}
where $D$ is the wire diffusion coefficient,  the parameter $\Gamma$ accounts for the inelastic scattering rate within the relaxation time approximation~\cite{Dynes_Tunneling_1984,Pekola_Limitations_2004,Pekola_Environment-Assisted_2010, Saira_Vanishing_2012}, and $\hat \tau_z$ is the third Pauli matrix in the Nambu space. 
The Green's function $\hat g$ is a matrix in this space with the form 
$\hat g=g\hat\tau_z+\hat f,$ 
being $\hat f$ the anomalous Green's function that is off-diagonal in Nambu space. 
The Usadel equation has to be complemented by the normalization condition $\hat g^2=\hat 1$, where $\hat 1$ is the unit matrix. 
The SN interfaces are modeled by proper boundary conditions. If the SN interfaces have large contact resistance $R_{SN}$ we use the Kupiyanov-Lukichev boundary conditions~\cite{kuprianov_influence_1988} $\left.\hat g\partial_x\hat g\right|_{x=\pm L/2}=\pm \frac{1}{2R_{SN}\mathcal{A}\sigma}\left[\hat g_\pm,\hat g\right]$,
where  $\mathcal{A}$ is  cross sectional area of the SN interface, $\sigma$ the  conductance of the N-wire and $\hat g_\pm$ are the Green's functions of the left and right S electrodes defined as 
$\hat g_\pm=g_s\hat \tau_3+f_s\left[ \cos(\varphi/2) i\tau_2\pm \sin(\varphi/2)i\tau_1\right]$.
Here, 
$g_s(\varepsilon)=\frac{\varepsilon+i\Gamma}{\sqrt{(\varepsilon+ i\Gamma)^2-\Delta^2}}$,  
$f_s(\varepsilon)=\frac{\Delta}{\sqrt{(\varepsilon+i\Gamma)^2-\Delta^2}}$,  
$\Delta$ is the BCS temperature-dependent energy gap with critical temperature $T_c=(1.764k_B)^{-1}\Delta_0$, $\Delta_0$ is the zero-temperature order parameter and $k_B$ is the Boltzmann constant. Furthermore, we neglect the suppression of the ring order parameter  at the SN interfaces due to \emph{inverse} proximity effect. This latter can be made negligible by making the wire cross section much smaller than that of the S loop~\cite{le_sueur_Phase_2008,hammer_Density_2007,Giazotto_Superconducting_2010}. If the SN contact resistance is negligibly small one imposes the continuity of the Green's functions at the interface.

We first focus on the \emph{short} junction limit (i.e., $E_{Th}=\hbar D/L^2\gg\Delta_0$, where  $E_{Th}$ is the SNS junction Thouless energy).  In this case the proximity effect can be maximized, and the performance of the nanovalve enhanced. 
This limit can be practically met with a copper N wire ($D = 0.01~{\rm m}^2/$s) of 100~nm length and an aluminum S ring ($\Delta_0=200\,\mu$eV) for which $E_{Th} = 3.3~ \Delta_0$. 
In such a case, the DoS in N can be obtained analytically: 
\begin{equation}
\label{shortDOS}
\mathcal{N}_{wire}= \left|
{\rm Re} 
\left[
\frac{\varepsilon-2i\varepsilon_bg_s}{\sqrt{(\varepsilon-2i\varepsilon_b g_s)^2+[2\varepsilon_b f_s\cos (\frac{\pi \Phi}{\Phi_0})]^2}}
\right]
\right|,
\end{equation}
where $\varepsilon_b= \hbar D/(2LR_{SN}\mathcal{A}\sigma)= \gamma E_{Th}$, and $\gamma\equiv R_{N}/R_{SN}$ quantifies the transmissivity of the SN contact~\cite{Giazotto_Quantum_2013}.

Figure~\ref{fig1}(b) and (c) show $\mathcal{N}_{wire}$ vs energy $\varepsilon$  for different $\gamma$ values at $\Phi=0$. 
A clear energy gap $\Delta_0$ in the DoS is visible even for quite low interface transmissivity (down to $\gamma \sim 0.1$) while for smaller $\gamma$ we observe the suppression of the minigap due to the weak coupling with S. 
Moreover, the minigap damping is bounded by the generation of a secondary gap appearing for $\varepsilon \lesssim \Delta_0 $, similarly to what has been predicted in Ref.~\cite{Levchenko_Crossover_2008, Reutlinger_Smile_2014}.
In the following, unless differently stated, we will set $\Gamma = 10^{-4}\,\Delta_0$ and $\gamma = 10$, as estimated from realistic values of transparent Al/Cu SN interfaces~\cite{Courtois_Origin_2008}.
For this trasmissivity the external magnetic flux $\Phi$ can efficiently modulate the minigap of the DoS in accordance to Eq. (\ref{shortDOS}). Figure~\ref{fig1}(d) and (e) show this modulation, where a clear quenching of the minigap is visible for $|\Phi| =  \Phi_0 /2$.  
This DoS modulation is the working principle of the present device and allows heat transport when the wire is in the normal state (i.e., for $|\Phi| =  \Phi_0 /2$) whereas it provides large thermal isolation when it is in the superconducting one. 
Full control over the heat current flowing through the device is one of the peculiar properties of this nanovalve that, differently from the other phase-coherent thermal modulators~\cite{Giazotto_Josephson_2012, Giazotto_Phase-controlled_2012,Martinez-Perez_Fully_2013,Martinez-Perez_quantum_2014,Eom_Phase_1998, Dikin_Low-temperature_2001,Bezuglyi_Heat_2003}, allows an almost complete quenching of the heat flow. 
\begin{figure}[t!]
\includegraphics[width=\columnwidth]{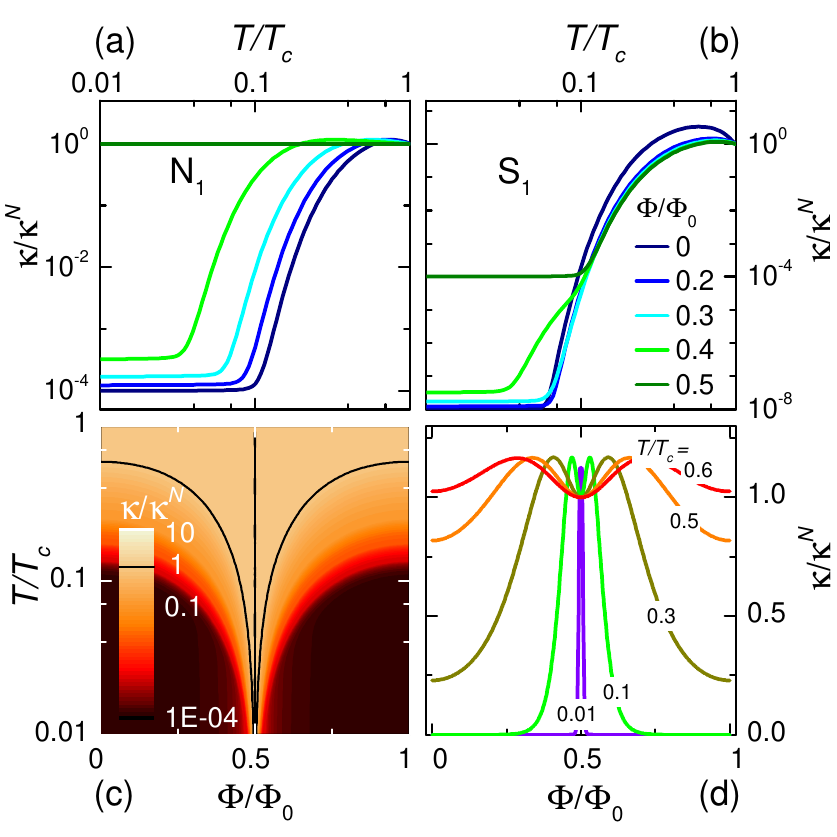}
\caption{\label{fig2}
Comparison between the  nanovalve thermal conductance $\kappa$ vs temperature $T$ calculated for a N$_1$, (a), and a S$_1$, (b), electrode. 
(c) Color plot of $\kappa$ calculated in (a) vs $\Phi$ and $T$ and (d) cross sections of it at selected temperatures $T$. 
}
\end{figure}
The thermal current ($J$) flowing from the SN ring to the N$_1$(S$_1$) electrode [see Fig.~\ref{fig1}(a)] can be written as~\cite{Giazotto_Opportunities_2006}
$J=\frac{2}{e^2R_T}\int_{0}^{\infty} d\varepsilon \varepsilon \mathcal{N}_{wire}(\varepsilon,\Phi)\mathcal{N}(\varepsilon)\mathcal{F}(\varepsilon)$,
where $\mathcal{N}(\varepsilon)=1$ for N$_1$, $\mathcal{N}(\varepsilon)=|{\rm Re}[(\varepsilon+i\Gamma)/\sqrt{(\varepsilon+i\Gamma)^2-\Delta^2}|$ for S$_1$,
$\mathcal{F}(\varepsilon)=[f_0(\varepsilon,T_L)-f_0(\varepsilon,T_R)]$,
$f_0(\varepsilon,T_{L,R})=[1+\mbox{exp}(\varepsilon/k_BT_{L,R})]^{-1}$ is the Fermi-Dirac distribution function, end $e$ is the electron charge. 
When the temperature difference between the SN ring and the N$_1$(S$_1$) electrode is small, $\delta T\equiv T_L-T_R \ll T\equiv (T_L+T_R)/2$, we  can define the thermal conductance  in the linear-response regime, $\kappa \equiv J/ \delta T$, which can be written as 
\begin{equation}
\kappa =\frac{\alpha}{T^2}\int_{0}^{\infty} d\varepsilon \varepsilon^2 \mathcal{N}_{wire}(\varepsilon,\Phi)\mathcal{N}(\varepsilon)\text{sech}^2\left(\frac{\varepsilon}{2k_B T}\right),
\label{conductance}
\end{equation} 
where $\alpha=(2e^2k_BR_T)^{-1}$.
For $T>T_c$, Eq. (\ref{conductance}) reduces to $\kappa^N=\mathcal{L}_0T/R_T$ (Wiedemann-Franz law) where $\mathcal{L}_0=\pi^2 k_B^2/3e^2$ is the Lorenz number.

Figure~\ref{fig2}(a) and (b) display the behavior of $\kappa (T)$ calculated for a few values of $\Phi$, for a normal ($N_1$) and a superconducting (S$_1$) electrode, respectively.
At low temperature ($T\lesssim0.1\,T_c$) a large suppression of $\kappa$ can be achieved for $\Phi=0$ (i.e., down to $10^{-4} \kappa ^N$) due to the presence of a S-like DoS in the wire which leads to a reduction of quasiparticles available for thermal transport, and which is only limited by the finite value of $\Gamma$. 
At fixed $T$, $\kappa$ then gradually increases, as displayed in Fig.~\ref{fig2}(c) and (d), eventually coinciding with  $\kappa^N$ by closing the minigap at $\Phi_0/2$. 
We emphasize that although a heat valve effect could be achieved as well by replacing the N$_1$ electrode with a superconductor S$_1$ [see Fig~\ref{fig2}(b)], the performance of the resulting structure, apart a stronger quenching of $\kappa$ at low $T$, worsen owing to the presence of the energy gap in S$_1$ which severely limits the heat current flow. 
For this reason in the following we will focus on the properties of the nanovalve implemented with a N$_1$ electrode. 
In order to obtain large $\kappa$ modulations a tunnel contact between the N wire and N$_1$ is crucial. The presence of a \emph{clean} metallic contact indeed leads to a drastic degradation of the superconducting-like properties of the N-wire due to \emph{inverse} proximity effect \cite{Bezuglyi_Heat_2003}. 
\begin{figure}[t!]
\includegraphics[width=\columnwidth]{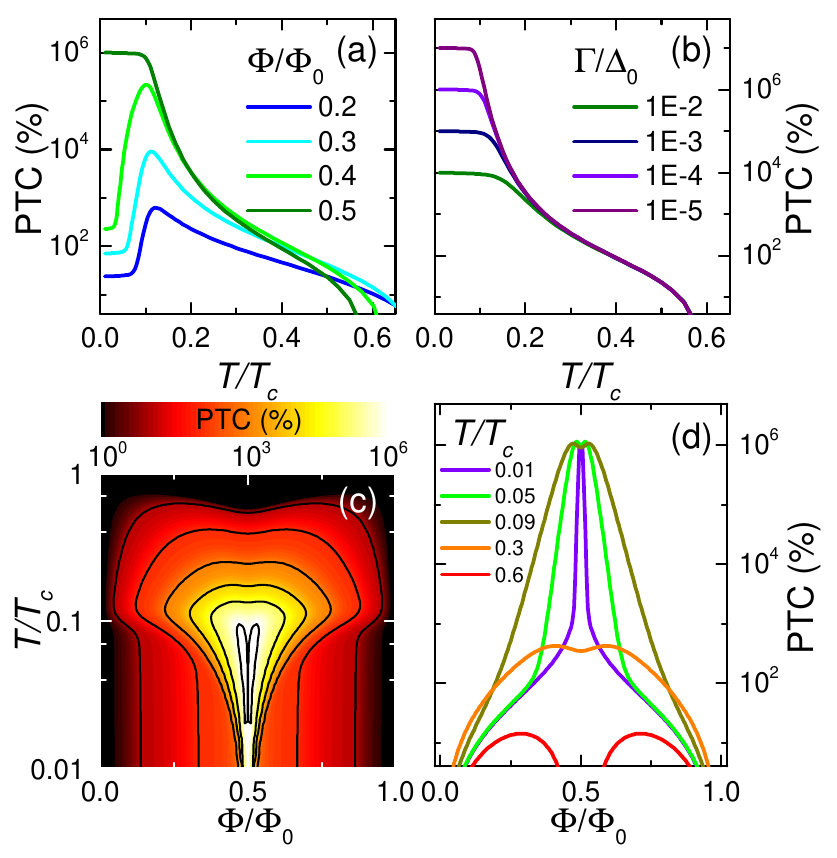}
\caption{\label{fig3}
Phase-dependent thermal conductance ratio (PTC) vs $T$ calculated for few values of $\Phi$, (a), and for a few values of $\Gamma$ at $\Phi= \Phi_0 /2$, (b). 
(c) PTC vs $\Phi$ and $T$. (d) Representative cross sections of panel (c) at selected  temperatures $T$.
}
\end{figure}

The heat valve efficiency of this setup can be quantified through the phase-dependent thermal conductance ratio (PTC) defined as: $\textrm{PTC}(T,\Phi)=[\kappa(T,\Phi)-\kappa(T,0)]/\kappa(T,0)$.
As can be noticed in Fig.~\ref{fig3}(a) and (c), where the PTC ratio is calculated vs $T$ and $\Phi$, respectively, the nanovalve can be highly efficient at temperatures below $\sim T_c/2$ with a PTC exceeding 100\% for $T< 0.2\,T_c$, and saturating to $ \Delta_0 / \Gamma$ at $T\lesssim 0.1\,T_c$ for $\Phi = \Phi_0 / 2$, as shown in Fig.~\ref{fig3}(b).
Moreover, at low temperature, the thermal valve is more sensitive to the magnetic flux $\Phi$, as demonstrated by the sharp open/close transitions appearing around $\sim\Phi_0/2$ in Fig.~\ref{fig3}(d). 

\begin{figure}[t!]
\includegraphics[width=\columnwidth]{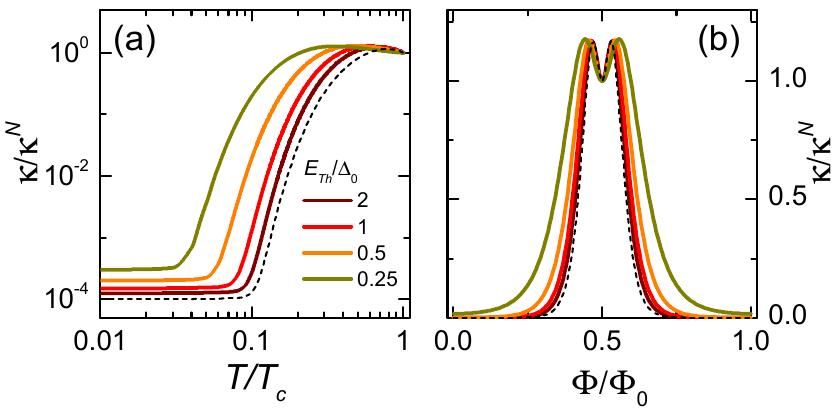}
\caption{\label{fig4}
Comparison between the nanovalve thermal conductance $\kappa$ calculated from the solution of the Usadel equations for different $E_{Th}$ (continuous lines) and in the \emph{short} junction limit (dashed lines) vs $T$ for $\Phi = 0 $, (a), and vs $\Phi$ for $T=0.1\,T_c$, (b).
}
\end{figure}

All the above results have been obtained from Eq. (\ref{shortDOS}), which is valid in the limit of a short N-wire. 
In the case of an arbitrary length we have solved numerically the Usadel equation (\ref{Usadel}) in the N region\cite{Virtanen_Thermoelectric_2007, Virtanen_Quasiclassical_2009} to obtain the DoS in the middle of the wire and compute the thermal conductance $\kappa$ from Eq.~(\ref{conductance}). 
The solutions are shown in Fig.~\ref{fig4} where the behavior of $\kappa$ in $T$ and $\Phi$ is displayed for different $E_{Th}$. 
At large $E_{Th}$ the solutions converge to the analytical one (represented by the dashed lines) that well approximates $\kappa$ for $E_{Th} \gtrsim  \Delta_0 $. 
Moreover, the general trend of $\kappa$ in $T$ and $\Phi$ is reproduced also for longer wires ($E_{Th} < \Delta_0 $) with a damping of $\kappa$ still three order of magnitude smaller than $\kappa_N$ at low temperatures (i.e., for $T<~0.05\,T_c$), therefore ensuring full functionality of the thermal nanovalve even for $E_{Th} = 0.25\,\Delta_0 $. 
\begin{figure}[t!]
\includegraphics[width=\columnwidth]{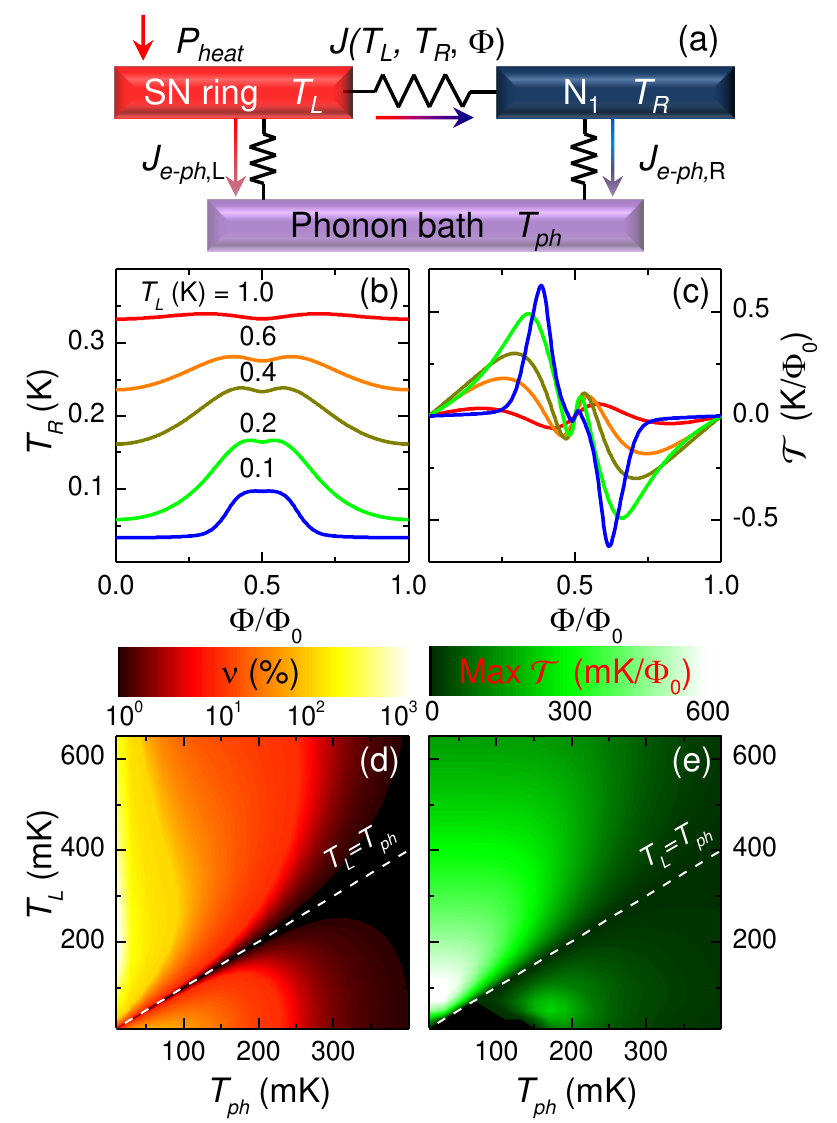}
\caption{\label{fig5} 
(a) Sketch of the thermal model accounting for heat transport in the proximity nanovalve. $J_{e-ph,L(R)}$ represents the heat current flowing between quasiparticles and lattice phonons residing at $T_{ph}$ in the left(right) electrode, whereas $P_{heat}$ denotes the power injected into the SN ring in order to impose a quasiparticle temperature $T_L$. The arrows indicate the direction of heat currents for $T_L > T_R > T_{ph}$.
(b) Temperature $T_R$ vs $\Phi$ calculated  for selected  $T_L$ assuming  $T_{ph} = 20$~mK.
(c) Flux-to-temperature transfer function $\mathcal{T}$ vs $\Phi$ calculated  from the data in panel (b). 
Color plots of the thermal visibility $\nu$, (d),  and the maximal transfer function, (e), achievable with the proximity nanovalve vs $T_{ph}$ and $T_L$. 
}
\end{figure}

According to the above conditions the experimental realization of our thermal nanovalve can be easily achieved with conventional metals and standard lithographic techniques~\cite{Giazotto_Superconducting_2010, Meschke_Tunnel_2011, Jabdaraghi_Non-hysteretic_2014, Ronzani_Highly-sensitive_2014}. 
Superconducting tunnel junctions additionally coupled to the SN ring and the N$_1$ electrode, serving either as heaters or  thermometers, allow to change and monitor the quasiparticle temperature on both sides of the structure~\cite{Giazotto_Opportunities_2006}. We neglect here  the contribution to thermal transport through these probes as they can provide nearly-ideal thermal isolation of the nanovalve electrodes. 

Figure \ref{fig5}(a) shows the relevant model accounting for thermal transport in the device. Upon intentionally heating the SN ring at $T_L$ the steady-state electron temperature $T_R$ will depend on the heat exchange mechanisms occurring in N$_1$. 
Below $\sim 1$ K, the energy relaxation mechanism in N$_1$ stems mainly from electron-acoustic phonon interaction \cite{Giazotto_Opportunities_2006}, 
$J_{e-ph,R}(T_R,T_{ph})=\Sigma \mathcal{V}(T_R^n-T_{ph}^n)$, 
which allows heat exchange between quasiparticle and lattice phonons residing at $T_{ph}$.
Above, $\Sigma$ is the material-dependent electron-phonon coupling constant, $\mathcal{V}$ is the volume, and $n$ is the characteristic exponent of the N$_1$ metal. 
In the model we neglect thermal transport mediated by photons \cite{schmidt_Photon-mediated_2004,meschke_Single-mode_2006,pascal_Circuit_2011} as well as pure heat conduction by phonons \cite{maki_entropy_1965}.
For any given $T_{ph}$ and $T_L$, the steady-state $T_R(\Phi)$ is then obtained by solving the thermal balance equation $-J(T_L,T_R,\Phi)+J_{e-ph,R}(T_R,T_{ph})=0$.
For the following calculations we assume an aluminum (Al) ring with $\Delta_0=200\,\mu$eV, $\Gamma=10^{-4} \Delta_0$, $R_T=100\,\text{k}\Omega$, $\mathcal{V}=2\times 10^{-20}$ m$^3$, 
$n=6$ and $\Sigma=4\times 10^9$ WK$^{-6}$m$^{-3}$ as appropriate for an AlMn N$_1$ electrode~\cite{Martinez-Perez_Coherent_2014,Martinez-Perez_quantum_2014}. 
Furthermore, we assume the SNS junction to be in the \emph{short} limit which, according to the above discussions, properly describes the framework of a realistic nanovalve~\cite{Meschke_Tunnel_2011,Jabdaraghi_Non-hysteretic_2014, Ronzani_Highly-sensitive_2014}.

Phase-dependent control of thermal current through the nanovalve is demonstrated by the strong modulation of $T_R(\Phi)$ displayed in Fig.~\ref{fig5}(b) for different values of $T_L$ at $T_{ph}= 20$~mK. 
In particular, it can exceed $100$~mK at $T_{L} = 200$~mK.
The high response of the heat valve, quantified by the flux-to-heat current transfer coefficient ($\mathcal{T} \equiv \partial J/\partial \Phi$)~\cite{Giazotto_Phase-controlled_2012,Giazotto_Josephson_2012}, is demonstrated in Fig.~\ref{fig5}(c) where $\mathcal{T}$ is plotted for the same $T_L$ values of panel (b). 
In particular, $\mathcal{T}$ obtains values as large as $\sim 500$~mK$/\Phi_0$ at 20~mK and for $T_L\lesssim 0.1\,T_c$, and it keeps increasing at lower $T_L$, as shown in Fig.~\ref{fig5}(e).
Moreover, to quantify the visibility of temperature $T_R$ modulation induced by the magnetic flux we define  the parameter
$\nu = \delta T_R/T_{ph}$, where $\delta T_R= \text{max}[T_R(\Phi)]-\text{min}[T_R(\Phi)]$.
A full characterization of $\nu$ as a function of $T_L$ and $T_{ph}$ is shown in Fig.~\ref{fig5}(d). 
According to the calculations, a sizeable  $T_R$ modulation is still visible ($\nu > 10 \%$) up to 300~mK of bath temperature and, differing from the SQUID-based thermal modulator~\cite{Giazotto_Phase-controlled_2012}, the visibility is strongly enhanced at lower $T_{ph}$ exceeding 1000\% at 20~mK. 
Notably, the proximity nanovalve demonstrates good performance also for the temperature regime where $T_L<T_{ph}$, as shown by the high efficiency and large transfer functions visible in the regions below the white dashed lines of panels (d) and (e). 

In summary, we have proposed a thermal nanovalve based on the SQUIPT technology which is able to efficiently conduct or isolate heat depending on the magnitude of an applied  magnetic flux. Under experimentally accessible conditions the device can provide full phase control of the thermal conductance, which is unique at cryogenic temperatures. 
Standard nanolithographic techniques and conventional metals provide a straightforward route towards the implementation of this thermal nanovalve. 

We acknowledge P. Virtanen for fruitful discussions. The work of E.S. and F.G. was partially funded by the European Research Council under the European Union's Seventh Framework Programme (FP7/2007-2013)/ERC grant agreement No. 615187-COMANCHE, and by the Marie Curie Initial Training Action (ITN) Q-NET 264034. The work of F.S.B was supported by the Spanish Ministry of Economy and Competitiveness under Project No. FIS2011-28851-C02- 02 the Basque Government under UPV/EHU Project No. IT-756-13.
\bibliographystyle{apsrev4-1}
 \bibliography{Biblio}

\end{document}